\documentclass[aps,prl,twocolumn,groupedaddress]{revtex4}

\usepackage{graphicx} 
\usepackage{amsfonts}
\usepackage{amsmath}
\usepackage{amssymb}
\usepackage{amsfonts}
\usepackage{graphicx}

\begin{document}

\title{Bipolar High Field Excitations in Co/Cu/Co Nanopillars}
\author{B. \"{O}zyilmaz and A. D. Kent}
\affiliation{Department of Physics, New York University, New York, NY 10003, USA}
\author{M. J. Rooks and J. Z. Sun}
\affiliation{IBM T. J. Watson Research Center, P.O. Box 218, Yorktown
Heights, NY 10598, USA}

\begin{abstract}
Current-induced magnetic excitations in Co/Cu/Co bilayer
nanopillars ($\sim$50 nm in diameter) have been studied
experimentally at low temperatures for large applied fields
perpendicular to the layers. At sufficiently high current densities excitations, which lead to
a  decrease in differential resistance, are observed for both current polarities. Such
bipolar excitations are not expected in a single domain model of spin-transfer. We propose
that at high current densities strong asymmetries in the longitudinal spin accumulation cause
spin-wave instabilities transverse to the current direction in bilayer samples, similar to
those we have reported for single magnetic layer junctions.
\end{abstract}
\date{June 25, 2004}




\pacs{75.60.Jk, 75.30.Ds, 75.75.+a}

\maketitle
Recent experiments have confirmed the seminal predictions
by Slonczewski \cite{Slonczewski1996} and Berger \cite{Berger1996}, that
a magnet acting as a spin-filter on a traversing current can experience
a net torque, known as a spin-transfer torque. Spin-current induced magnetization
reversal \cite{Katine2000,Grollier2003,Ozyilmaz2003}, excitations \cite{Tsoi1998,Urazhdin2003,Fabian2003}
and magnetization precession  \cite{Kiselev2003,Rippard2004} have been directly observed in magnetic nanostructures.
These experimental studies of spin-transfer have focused on spin valve type
structures of ferromagnet/normal metal/ferromagnet layers, in which the layers may be non-collinear and
one of the layers is thicker than the other. This thicker layer serves as a reference layer that sets
up a spin-polarized current with a component of angular momentum transverse to the thin layer's magnetization.
In all these experiments current induced excitations have been observed for only one polarity of the
current, nominally because of the asymmetry in the layer structure. This observation was considered to be
unmistakable evidence for physics associated with a spin-transfer torque--as opposed to the effects
of charge-current induced magnetic fields. In addition, the lowest resistance state was always considered to be the
static state of parallel magnetic alignment.

Here we report studies of current-induced excitations of the
magnetization in Co/Cu/Co bilayer nanopillar junctions.
Experiments were performed at $T=4.2$ K in high magnetic fields
($H> 4\pi M$) in the field perpendicular to the plane geometry.
For sufficiently large current densities we observe anomalies in
$dV/dI$ \textit{independent} of current polarity, which decrease
the junction resistance. The bipolarity of the excitations and the
\textit{decrease} in resistance cannot be understood in terms of
spin-transfer torque induced single domain dynamics. These results
show that high current densities can induce excitations of the
magnetization independent of current polarity and relative
alignment of the magnetizations of the two magnetic layers. From
detailed I-V measurements we construct a phase diagram that shows
the conditions under which such excitations occur. In addition,
the results illustrate that at high currents the nanopillar
resistance can be lower than that of a state of parallel magnetic
alignment. We suggest that structural asymmetries in nanopillar
junctions lead to a longitudinal spin-accumulation pattern that
provides a new source for spin-transfer torque induced
magnetization dynamics. We have recently shown that the
longitudinal spin accumulation has an important influence on the
magnetization excitations in asymmetric single layer junctions
\cite{Ozyilmaz2004}. Here we show that similar physics is relevant
to the more typically studied bilayer devices that consist of a
`fixed' thick magnetic layer and a thin `free' magnetic layer.

\begin{figure}[t]
\includegraphics[width=0.48\textwidth]{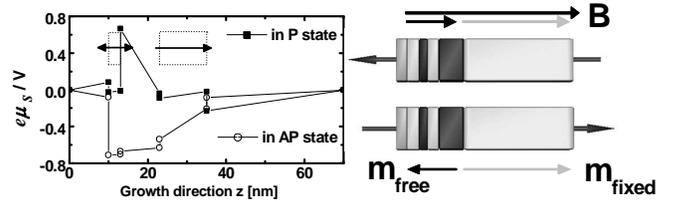}
\caption{Right: Structure of nanopillar device. The layer structure from left to right is Pt/Cu/Co/Cu/Co/Cu. The field, electron flow and magnetization directions are indicated. Calculated spin-accumulation pattern in the P state for negative currents and in the  AP state for positive currents, based on a two channel model, including interface resistances.
}
\label{Nanopillar}
\end{figure}
\begin{figure}[t]
\includegraphics[width=0.48\textwidth]{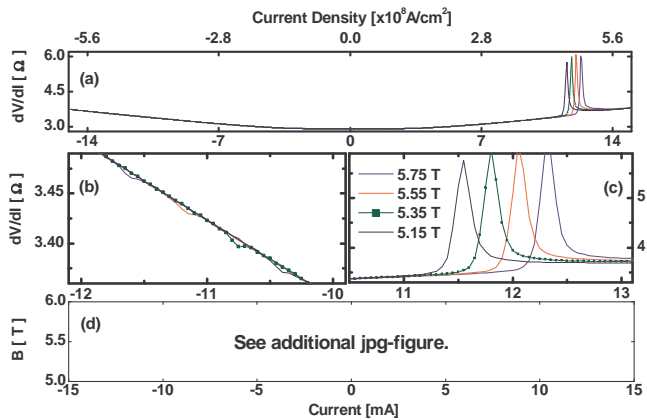}
\caption{(a-c) Differential resistance versus current with large applied perpendicular fields. (b) Dips in $dV/dI$ are observed at negative current bias under conditions for which instabilities are not expected in a single domain model. (c) The peak in $dV/dI$ at positive current and the change in differential resistance after the peak is due to switching into an AP state. (d) Color plot showing $d^2V/dI^2$ versus current and applied field (red is high and blue is low). The dispersion of the dip in $dV/dI$ at negative current with increasing field is clearly visible. This dip is a clear signature of nonuniform magnetic excitations.}
\label{PeaksDips}
\end{figure}

We study spin-transfer torques in devices fabricated by means of a
nano-stencil mask process \cite{Sun2002,Sun2003}. This approach
defines the lateral dimensions ($\sim 50$ nm$\times50$ nm) of the
junction prior to the growth of a $Pt 15$ nm$| Cu 10 $ nm$|Co 3 $
nm$|Cu 10 $ nm$|Co 12 $ nm$|Cu 300$ nm multilayer. The layer
structure is illustrated schematically in Fig. \ref{Nanopillar}. Six such
junctions were studied in detail and representative data on one
junction is presented in this paper. Transport measurements were
conducted using a four probe geometry, where the differential
resistance $dV/dI$ was measured by means of a phase sensitive
lock-in technique with a $100 \mu$A modulation current at $f=873$
Hz added to a dc bias current.  
Positive current is defined such that the electrons flow from the
thin ferromagnetic layer to the thick ferromagnetic
layer. Fig. \ref{PeaksDips} (a) shows typical measurements of
$dV/dI$ versus $I$ in large applied fields ($B>>B_{demag} \sim 1
$ T). Current induced excitations (peaks in $dV/dI$) appear to
occur only at positive current bias. However, a more careful look
at the current sweeps reveals the presence of excitations even at
negative currents (Fig. \ref{PeaksDips} (b)). Here we observe
anomalies in $dV/dI$ in the form of $\emph{dips}$. These dips
correspond to decreases in the differential resistance of about
$0.5$\%. These excitations shift to higher currents as we increase
the applied field. They are best distinguished from the parabolic
background resistance by plotting the second derivative on a color
scale as a function of current bias and applied field. From this
color plot (Fig. \ref{PeaksDips} (d)) we see that at both polarities
the excitations depend linearly on the applied field.

Peaks in $dV/dI$ at positive currents are well understood. Earlier
work \cite{Ozyilmaz2003} has shown that their position indicates
the critical current, $I_{crit}$,  necessary to switch the free layer
magnetization into the high resistance state of anti-parallel (AP)
alignment. This interpretation is further supported by recent high frequency
noise experiments in the field perpendicular to the plane geometry \cite{Kiselev2004}.
Excitations at negative currents are
unexpected. In the parallel configuration (P) negative currents are
expected to suppress any deviation of the free layer from
parallel alignment with the fixed layer. In particular, in large applied fields, the
layer magnetizations should therefore remain in the P state. In
addition, we observe dips instead of peaks in $dV/dI$, indicating
that excitations at negative currents decrease the junction
resistance. However, within a single domain model the giant
magnetoresistance effect (GMR) should lead to an increase of the
junction resistance whenever the layer magnetizations deviate from
parallel alignment.

\begin{figure}[t]
\includegraphics[width=0.48\textwidth]{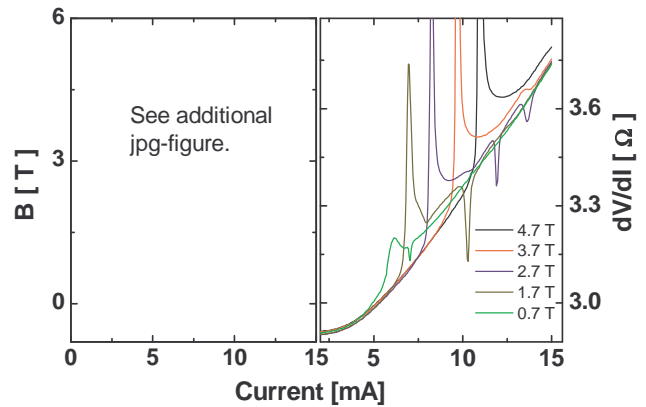}
\caption{$dV/dI$ versus current at large positive current. Color plot showing $d^2V/dI^2$ versus current and applied field. Dips in differential resistance after the main peak in $dV/dI$ (labeled B) are visible and correspond to boundary C. }
\label{PosBias}
\end{figure}

Further, at positive currents in a single domain model in which the thick layer magnetization
remains fixed there are no further excitations once the AP state is achieved, i.e.,
once $I> I_{crit}$, after the main peak in $dV/dI$.
However, there is structure in $dV/dI$ beyond the main
peak, again in the form of dips in $dV/dI$. The results are shown in Fig. \ref{PosBias}.
Here we plot the differential resistance as a function of current for selected applied fields,
$0.7 $ T$ < B < 4.7 $ T. We observe both peaks \emph{and} dips in
$dV/dI$. However, at fields $B > 1$ T, dips occur only for
$I > I_{crit}(H)$. Also, most current sweep traces show multiple
dips in $dV/dI$. The field dependence of these excitations is best
seen when the second derivative $d^{2}V/dI^{2}$ is plotted on a
color scale as a function of current bias and applied field (Fig.
\ref{PosBias} (a)). Such a plot reveals two boundaries, which can be best
distinguished at fields $B> 1.5 $ T. The first boundary (B)
represents the currents $I_{crit}(H)$ at which the free layer
switches into the AP state. Note that in many samples we also observe
additional peaks in $dV/dI$ for $I<I_{crit}(H)$. These peaks
coincide with small upward jumps of the junction resistance (not
shown), which we associate with the onset of current-induced
magnetization precession \cite{Kiselev2004}.

At higher currents the color plot reveals a second boundary (C). The latter
marks the critical currents $I_{+ }(H)$ above
which we observe dips in $dV/dI$. The current bias and field
dependence of these additional excitations is non-trivial. They are
best described by first considering cuts parallel to the current
axis of the color plots and then cuts parallel to the field axis.
In the first case the applied field is constant. Now as we
increase the current we cross several branches corresponding to
distinct excitations. At each of these crossings we observe dips
in $dV/dI$. From cuts parallel to the field axis (constant current
bias), we see that each excitation exists only in a very narrow
field range, i.e. they have a weak dependence on magnetic field. Note
that this is similar to the field dependence of $I_{crit}$ at
fields ($B<B_{demag} \sim 1$ T).
Also here the excitations shift
to lower currents as we increase the applied field. In addition,
different branches of excitations are separated by narrow stripes
of high resistance regions. We suspect that these gaps reflect the quantization of transverse
spin-wave modes in these small elements.

\begin{figure}[t]
\includegraphics[width=0.48\textwidth]{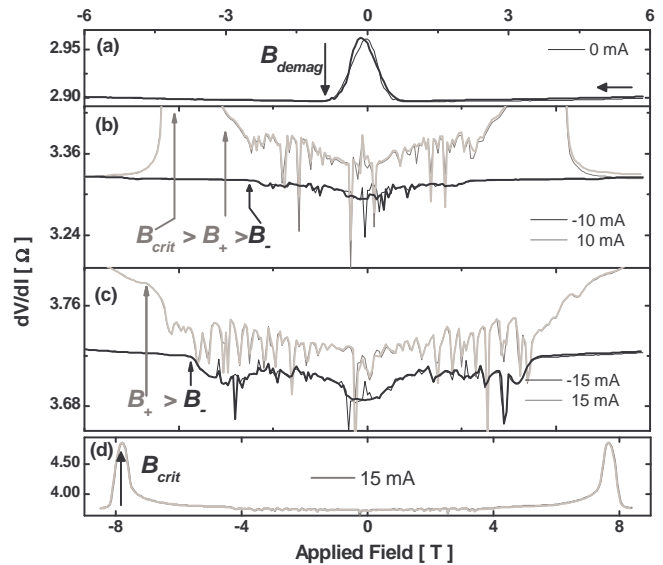}
\caption{$dV/dI$ versus applied field at constant current bias. (a) $I=0$, $dV/dI$ decreases as the layer magnetizations align with the applied field. Above the demagnetization field no excitations are observed. (b-c) I= $\pm 10 $ mA and $\pm 15 $ mA. Excitations are observed at large fields ($B>B_{demag}$) for both positive and negative currents, as described in the text. (d) $I =15$ mA. The peak at $B_{crit}$ marks the switching of the layers to a P state.}
\label{MR}
\end{figure}

The bipolarity of these high field, high current excitations
(i.e., the dips in differential resistance) can also be seen in
field sweeps at fixed current bias. We show examples of such
measurements in Fig. \ref{MR}. A field sweep at zero dc bias is
shown in Fig. \ref{MR}(a), whereas Fig. \ref{MR}(b) and 4(c) shows
the MR at $I=\pm10mA$ and $I=\pm15mA$ respectively. As shown in
Fig. \ref{MR}(a) at zero dc bias excitations at fields
$B>B_{demag}( \sim 1$ T) are absent in the field traces. However,
high current densities lead to excitations even at fields
$B>B_{demag}$, \text{independent} of current polarity. These
current driven excitations vanish once the magnitude of the
applied field exceeds critical fields $|B_{+}|$ at positive
currents and $|B_{-}|$ at negative currents. At positive bias we
see the additional boundary at $|B(I)=B_{crit}|$ at which the
applied field switches the free layer magnetization back to the P
state (Fig. \ref{MR}(d). The main difference between $|B_{crit}|$
and $|B_{+}|$,$|B_{-}|$ is that the former leads to a
\textit{decrease} in resistance whereas the latter indicate the
point where the junction resistance \textit{increases}.

We now discuss an interpretation of these results in terms of
spin-wave instabilities that are expected in the presence of
strong asymmetries in the longitudinal spin-accumulation
\cite{Polianski2004,Stiles2004}. First, we note that dips in
$dV/dI$ at negative current bias have been also observed in pillar
junctions with only a single ferromagnetic layer. Here,
excitations are consequence of asymmetric leads
\cite{Ozyilmaz2004}, which induce an asymmetry in longitudinal
spin accumulation. The necessary condition for such instabilities
is that the current bias has to be such that the sum of the
longitudinal spin accumulation on either side of the ferromagnetic
layer, i.e. the net spin accumulation, is opposite the
magnetization direction \cite{Polianski2004,Stiles2004}.

We have modeled the spin accumulation pattern in our bilayer
junctions using the two current model, with the spin-dependent
bulk and interface resistances of Ref. \cite{Yang1995} in the
limit in which the spin-diffusion length is much larger than the layer
thicknesses ($\lambda_{sf} \rightarrow \infty$). 
Fig. \ref{Nanopillar} shows that in the P state at
negative current bias the spin-accumulation about the thick layer
is asymmetric; the net spin accumulation is opposite the
magnetization direction. According to the condition governing
spin-wave instabilities in single layers
\cite{Polianski2004,Stiles2004}, this accumulation pattern can
excite nonuniform spin-waves in the thick layer.

To explain the new region of excitations at currents beyond
$I_{+}(H)$, i.e. excitations at positive current bias in the AP
state, we also consider the spin accumulation in this case
(Fig. \ref{Nanopillar}, AP state graph). From this we see that the
switching of the free layer has an important effect on the spin
accumulation pattern at the fixed layer. The sign of the net spin
accumulation inverts as the system is switched by the current from
the P state into the AP state. Therefore, excitations of the fixed
layer now require a \emph{positive} current bias. This is in
agreement with the experimental observation. From this we conclude
that dips in $dV/dI$ at \textit{both} positive and negative
currents are caused by excitation of the thick magnetic layer.
While at positive currents these excitations could be associated
with uniform excitations of the fixed layer, the pattern of the
excitations matches well the nonuniform excitations found in
single layers \cite{Ozyilmaz2004}. A longitudinal spin
accumulation opposite the magnetization direction on both sides of
this layer seems to be the most likely cause for these excitations.
At first the situation would appear to be quite similar to that
for which excitations have already been observed in single layer
junctions \cite{Ozyilmaz2004}. However, there are some notable
distinctions. In single layer junctions the presence of an
asymmetric and hence a non-vanishing net longitudinal spin
accumulation at the ferromagnetic layer is caused by different top
and bottom non-magnetic leads. In bilayers, with a thick and thin
layer there is a built-in asymmetry. So in contrast to single
layer junctions a lead asymmetry, is not  necessary for current
induced instabilities. Another importance difference is that the
excitations in the single layer junctions rely on a feedback
mechanism between the layer magnetization and the
spin-accumulation in the adjacent non-magnetic layers. In bilayer
junctions the layer magnetization is biased with a longitudinal
spin accumulation set by the spin-dependent bulk and interface
conductivities, relative size and orientation of the two
ferromagnetic layers \cite{BrataasPC}. Spin diffusion along the
interface, which provides the feedback, can compete with the
latter, but is not necessary to produce excitations. Hence, single
layer and (collinear) bilayer junctions probe spin-dependent
transport and spin-current induced excitations of the
magnetization under distinct conditions.

The presence of a second layer in bilayer junctions allows a
determination of the nature of these excitations. For
example, the decrease in junction resistance allows one to distinguish
between nonuniform and uniform excitations. Uniform excitations
at negative bias (P state) should always produce an increase of junction
resistance because of GMR. Only nonuniform excitations can account for a
decrease of junction resistance.
This can be explained by considering the effect of spin
accumulation on the junction resistance. Any spin accumulation
between the ferromagnetic layers will increase the junction
resistance. Nonuniform excitations effectively reduce the amount
of spin accumulation in the spacer layer, because they mix the two
spin channels \cite{Polianski2004,BrataasPC}. Hence, the junction resistance
decreases. We believe that this also explains why it is easier to
observe these excitations in bilayer junctions in the AP state than
in bilayer junctions in the P state or for that matter in single
layer junctions. Comparing Fig. \ref{PeaksDips} we see that the spin accumulation
in the spacer layer is largest in the AP state. Therefore the
largest reduction in device resistance will occur when nonuniform
excitations take place in the AP state.

In conclusion we have shown that intrinsic asymmetries in bilayer
junctions lead to current induced instabilities for both current polarities.
These excitations lower the junction resistance suggesting that the lowest resistance state occurs for
a P state with nonuniform excitations in the `fixed' thick magnetic layer. The decrease
in resistance for negative current polarities, from a P  magnetic state, is strong
evidence for nonuniform excitations.
\begin{acknowledgments}
We acknowledge useful discussions with A. Brataas, P. Brouwer and M. Stiles. This research is supported by grants from NSF-FRG-DMS-0201439, NSF-DMR-0405620 and by ONR N0014-02-1-0995.
\end{acknowledgments}


\end{document}